# Photonic crystal waveguide intersection design based on Maxwell's fish-eye lens

M. M. Gilarlue[a], S. Hadi Badri[b], H. Rasooli Saghai[c,]*, J. Nourinia[a], Ch. Ghobadi[a]

[a] Department of Electrical Engineering, Urmia University, Urmia 57153, Iran {m.gilarlue, j.nourinia, ch.ghobadi}@urmia.ac.ir

[b] Department of Electrical Engineering, Azarshahr Branch, Islamic Azad University, Azarshahr, Iran

[c] Department of Electrical Engineering, Tabriz Branch, Islamic Azad University, Tabriz, Iran

Corresponding author.

* E-mail addresses: h_rasooli@iaut.ac.ir

## Abstract

The number of waveguides crossing an intersection increases with the development of complex photonic integrated circuits. Numerical simulations are presented to demonstrate that Maxwell's fish-eye (MFE) lens can be used as a multiband crossing medium. In previous designs of waveguide intersection, bends are needed before and after the intersection to adjust the crossing angle resulting in a larger footprint. The presented design incorporates the waveguide bends into the intersection which saves footprint. In this paper, *4×4* and *6×6* intersections based on ideal and graded photonic crystal (GPC) MFE lenses are investigated, where *4* and *6* waveguides intersect, respectively. The intersection based on ideal MFE lens partially covers the O, E, S, C, L, and U bands of optical communication, while the intersection based on GPC-MFE lens is optimized to cover the entire C-band. For *4×4* and *6×6* intersections based on GPC-MFE lens, crosstalk levels are below *-24dB* and *-18dB*, and the average insertion losses are *0.60dB* and *0.85dB* in the C-band with lenses' radii of *7×a* and *10×a*, respectively, where *a* is the lattice constant of the photonic crystal.

## 1. Introduction

Increasing complexity of photonic integrated circuits demands efficient guiding of optical signals through waveguide intersections. Numerous methods are introduced to design an optimized two-dimensional (2D) photonic crystal (PC) *2×2* waveguide intersection such as using a resonant cavity at the center of the intersection[1-3], topology optimization [4], Wannier basis design and optimization [5], utilizing the symmetric properties of the propagation modes of square-lattice PC waveguide [6]. To the best of our knowledge, the intersection of multiple waveguides has not been proposed for any of the mentioned methods. Furthermore, waveguide bends are needed in these methods to adjust the crossing angle of waveguides at the intersection. Designing a low-loss and broadband PC waveguide bend [7] and an arbitrary angle waveguide bend [8] could be challenging. Recently, an inverse-designed star-crossings has been introduced that supports the intersection of multiple waveguides using a time-consuming evolutionary algorithm as a design method [9]. However, waveguide bends still are needed before and after these star-crossings to adjust the angle of intersection. As shown in Fig. 1(a), implementing a conventional *4×4* intersection requires the design of six *2×2* intersections and eight waveguide bends which may require up to *256* square microns footprint to minimize the bending loss and crosstalk at intersections [9]. Recently, researchers have shown interest in gradient index (GRIN) lenses such as Luneburg [10, 11], MFE [12-14] and Eaton [10, 14] lenses. In the proposed design, MFE lens is used as a crossing medium incorporating the waveguide bends in itself. MFE lens is a type of GRIN lens that focuses the radiation of a point source on its surface to the diametrically opposite point of the lens [15, 16]. The refractive index of the MFE lens is defined as

$$n_{lens}(r) = \frac{2}{1+(r/R_{lens})^2} \quad , \quad (0 \leq r < R_{lens}) \qquad (1)$$

where $R_{lens}$ is the radius of the lens and r denotes the distance from the center of the lens. The intersection based on the MFE lens and its refractive index distribution are shown in Fig. 1(b). The generalized MFE lens formula can be used for different applications [17, 18], however, the refractive index profile given in Eq. (1) is appropriate for our design.

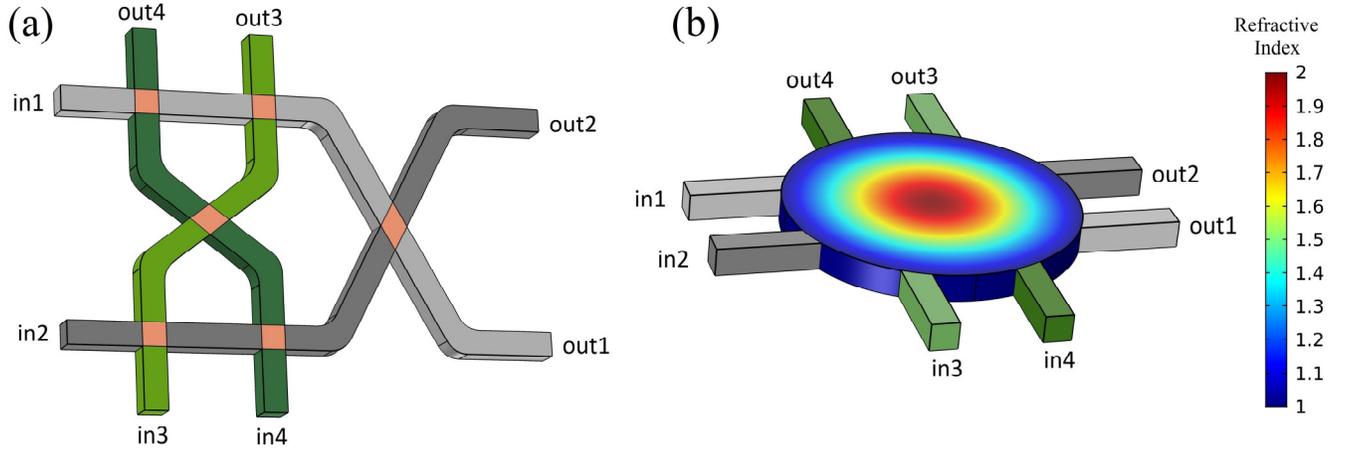

Fig. 1. Schematics of *4×4* intersection (a) implemented with *2×2* conventional intersections (b) implemented with Maxwell's fish-eye lens. Refractive index distribution of MFE lens is also shown.

## 2. Graded photonic crystal as GRIN medium

GPC in the metamaterial regime is a low-loss and broadband structure making it a viable candidate to implement GRIN medium [19]. GPC is used in implementation of different devices such as lenses [19-21], bends [22, 23], and optical couplers [24, 25]. In this paper, GPC is used to implement the MFE lens. The first step in implementing the GRIN medium with GPC is to grid the medium with cells of appropriate shape and size. Average value of the refractive index in each cell of the gridded medium is denoted as $n_{eff,ij}$ with $x_i$ and $y_j$ as coordinates of the cell's center. In the long-wavelength limit, the size of the unit cell is far smaller than the minimum wavelength of light. As a result, the medium can be treated as a homogeneous medium with an effective refractive index, which can be calculated with effective medium approximations (EMAs) theories such as Maxwell-Garnett approximation [26]. EMAs are used in implementation of different structures [27-31]. In contrast to the conventional definition of the transverse magnetic (TM) mode, the electric field intensity is parallel to the dielectric rods and perpendicular to the plane of wave propagation in PCs [32]. In TM mode, the medium can be regarded as homogeneous for even high volume fractions of rods. Consequently, the effective permittivity can be calculated by volume averaging theory [19, 33]

$$\varepsilon_{eff}^{TM} = f_{rod}\varepsilon_{rod} + (1-f_{rod})\varepsilon_{host} \qquad (2)$$

where $f_{rod}$ is the volume fraction of the rod in the cell, $\varepsilon_{eff}^{TM}$ is the effective permittivity of the cell for TM mode, $\varepsilon_{host}$ and $\varepsilon_{rod}$ are the permittivity of the host and rods, respectively. After rearranging, Eq. (2) can be expressed as

$$f_{rod} = \left(\varepsilon_{eff}^{TM} - \varepsilon_{host}\right) / \left(\varepsilon_{rod} - \varepsilon_{host}\right) \tag{3}$$

The volume fraction of cylindrical rods in the $ij$-th cell is $f_{rod,ij} = \pi r_{rod,ij}^2 / A_{ij}$, where $A_{ij}$ is the area of the $ij$-th cell, so the corresponding radius of the rod in the $ij$-th cell for TM mode is given by

$$r_{rod,ij} = \sqrt{\frac{A_{ij}(\varepsilon_{eff,ij}^{TM} - \varepsilon_{host})}{\pi(\varepsilon_{rod} - \varepsilon_{host})}} \tag{4}$$

For transverse electric (TE) mode, where the electric field is normal to the dielectric rods, the Maxwell-Garnett approximation is used [19]. Using this approximation, the effective permittivity for TE mode is given by

$$\varepsilon_{eff}^{TE} = \varepsilon_{host} + \frac{f_{rod}\varepsilon_{host}(\varepsilon_{rod} - \varepsilon_{host})}{\varepsilon_{host} + L(1 - f_{rod})(\varepsilon_{rod} - \varepsilon_{host})} \tag{5}$$

where $L$ is the depolarization factor depending on the shape of the rods (for cylindrical rods $L = 1/2$), $\varepsilon_{eff}^{TE}$ is the effective permittivity of the cell for TE mode. The corresponding radius of the rod in the $ij$-th cell for TE mode is given by

$$r_{rod,ij} = \sqrt{\frac{A_{ij}(\varepsilon_{eff,ij}^{TE} - \varepsilon_{host})(\varepsilon_{rod} + \varepsilon_{host})}{\pi(\varepsilon_{eff,ij}^{TE} + \varepsilon_{host})(\varepsilon_{rod} - \varepsilon_{host})}} \tag{6}$$

For the square cell $A_{ij} = a_{GPC}^2$ where $a_{GPC}$ is the lattice constant of the GPC.

## 3. Numerical simulation and discussion

In order to prove that the MFE lens can be used as a crossing medium, numerical simulations are performed using Comsol Multiphysics. The photonic crystal lattice is composed of cylindrical rods with refractive index of 3.4 in air host with the lattice constant of $a=540nm$ and radius of $r=0.185a$. A simple method for creating a TM mode waveguide in 2D photonic crystal consisting of dielectric rods in the square lattice is to remove a row of rods [34].

Perfectly matched layer (PML) and scattering boundary condition (SBC) are used in simulations to truncate the theoretically infinite extent to a finite computational domain.

Implementation of PML and SBC in PC waveguides has proven to be problematic. When only SBC is applied in PC simulations substantial reflection is observed from waveguide ends. This spurious reflection introduces large errors in scattering parameter calculations [35]. To overcome this issue, different methods are proposed such as terminating PC waveguide with conventional, distributed-Bragg-reflector [35], and PC-based PML [36]. In our simulations, as shown in Fig. 2, PC-based PML is used such that each port is truncated with an anisotropic PML domain which is surrounded with PC to eliminate crosstalk of evanescent wave in PML domain to adjacent ports. Due to the photonic bandgap of PC, the electromagnetic field is almost restricted to the waveguides and the intersection. So without introducing any spurious reflection, SBC is chosen for the remaining computational boundaries. The ports are used to calculate the scattering parameters in the frequency domain. The simulation domain of interest is denoted as PC and GPC-MFE lens which consists of four crossing waveguides and the GPC-MFE lens. It is worth mentioning that some rods of the main PC in the vicinity of the lens's surface are moved slightly.

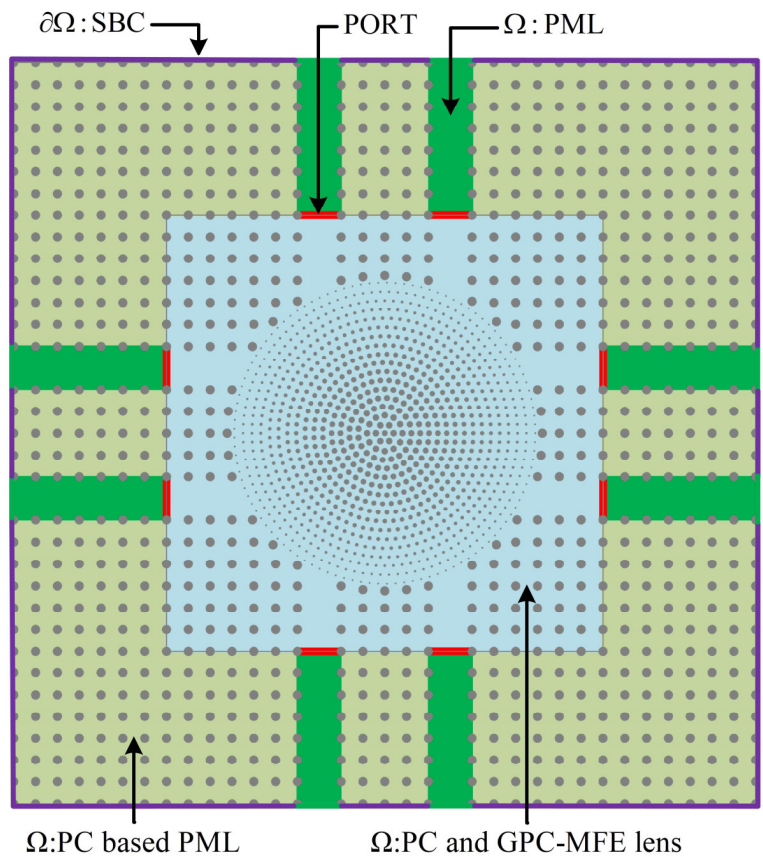

Fig. 2 Computational domains. PC-based PML is used to reduce spurious reflection from the ports. For all domains, the dielectric rods ($n=3.4$) in air are used in simulations.

The propagation of light from port *in1* to *out1* through *4×4* intersection based on the ideal MFE lens is illustrated at a wavelength of *1550nm* in Fig. 3(a). In this figure, power streams show the light propagation, moreover, extruded rods and the refractive index of the ideal lens is illustrated. The return loss of *15.0dB*, insertion loss of *0.58dB*, and crosstalk levels below *-21.4dB* are achieved, at this wavelength. The electric field distribution at *1550nm* is shown for the GPC-MFE lens based *4×4* intersection implemented with rods of varying sizes in Fig. 3(b). Although the overall performance of the ideal lens is better than the GPC-based lens, the scattering parameters of GPC-based lens are improved compared to the ideal lens, at this wavelength. The return and insertion losses are *23.3dB* and *0.42dB*, respectively, while crosstalk levels are less than *-29.4dB*. Moreover, the electric field's peaks and troughs of the *4×4* intersection are shown in Fig. 4, at a wavelength of *1550nm*. The simulated scattering parameters of the *4×4* intersection based on the ideal MFE lens are shown in Fig. 5 for optical communication bands. The intersection's performance is acceptable when the transmission is higher than *-3dB* and transmission's phase is linear. In Fig. 5, the wavelengths the intersection has the desired performance are highlighted in green. The ideal MFE lens based *4×4* intersection partially covers the O, E, S, C, L, and U bands.

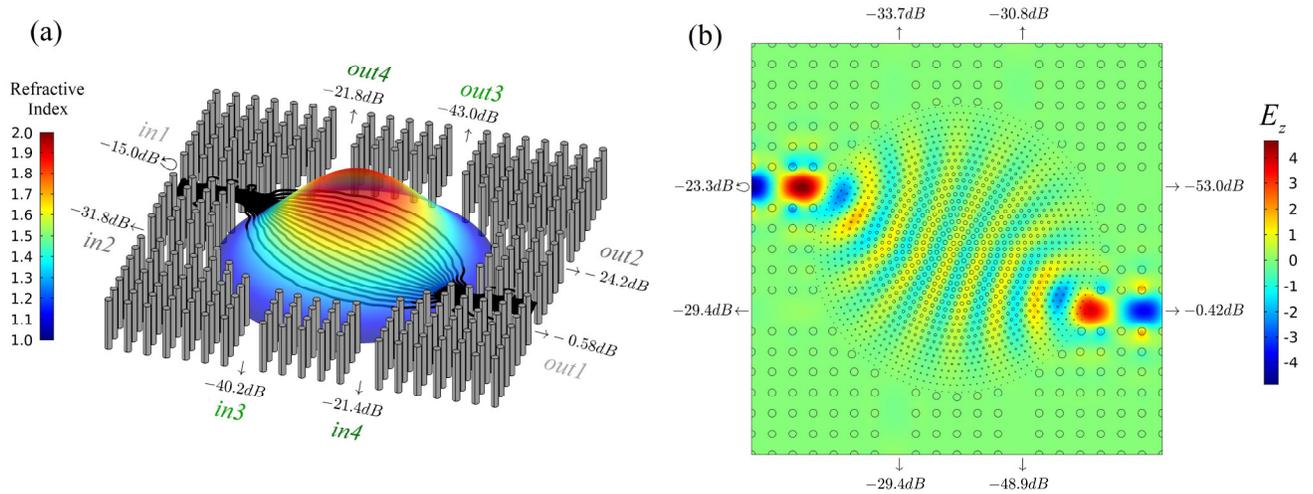

Fig. 3. Transmission, reflection, and crosstalk levels of 4×4 intersection based on the MFE lens at a wavelength of 1550nm. (a) Visualization of light propagation through the ideal MFE lens with power stream and extruded rods. (b) The electric field distribution of the MFE lens implemented with GPC.

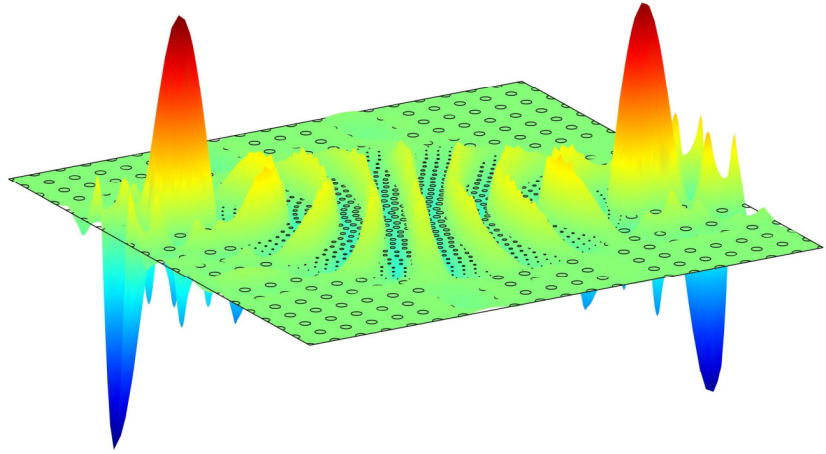

Fig. 4 The electric field distribution of *4×4* intersection based on GPC-MFE lens at *1550nm*.

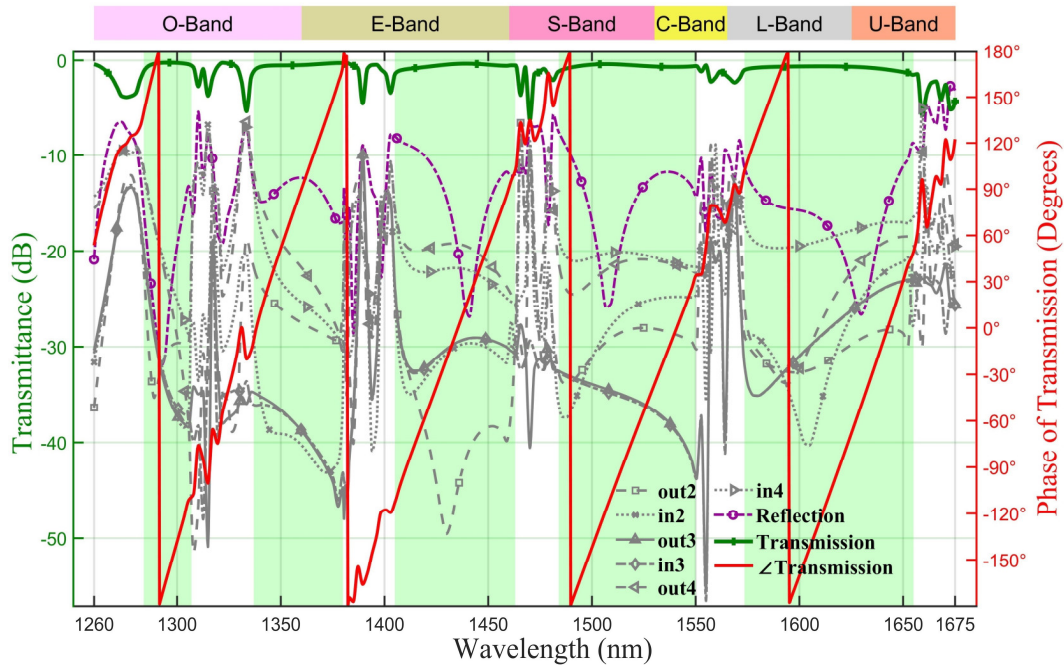

Fig. 5. Simulated transmission, reflection and crosstalk spectra of ideal MFE lens based *4×4* waveguide intersection of figure 3(a). The Wavelengths that the lens has the desired performance are highlighted in green.

In the implementation of GPC, the practical procedure is to use the same material as the main PC. There are two main approaches in the implementation of GPC. One method is to keep the rods' radii fixed and adjust the spacing between rods, i.e. cell size. The other method is to keep the cell size fixed and control the radii of rods. In the realization of the

*4×4* intersection with the lens radius of *7×a*, the latter method is used with optimum cell size to cover the entire C-band. The simulated scattering parameters of *4×4* intersection based on GPC-MFE lens are shown in Fig. 6. The crosstalk levels are below *-24dB* and the average insertion loss is *0.60dB* in the C-band. In this band, the transmission is almost flat and its phase is linear minimizing dispersion and pulse broadening. It is worth noting that in Fig. 5 and 6, some narrower transmission bands are ignored for both ideal MFE and GPC-MFE lenses. The transmission spectra of the *4×4* intersections based on the ideal MFE and GPC-MFE lenses are compared in Fig. 7 where green and blue horizontal bars specify the wavelengths with the acceptable transmission, respectively. In addition, the bandwidth of each transmission band is determined in this figure. Transmission bands of the ideal MFE based intersection have wider bandwidth compared to the GPC-MFE based intersection. On the other hand, the GPC-MFE based *4×4* intersection is optimized to cover the entire C and U bands.

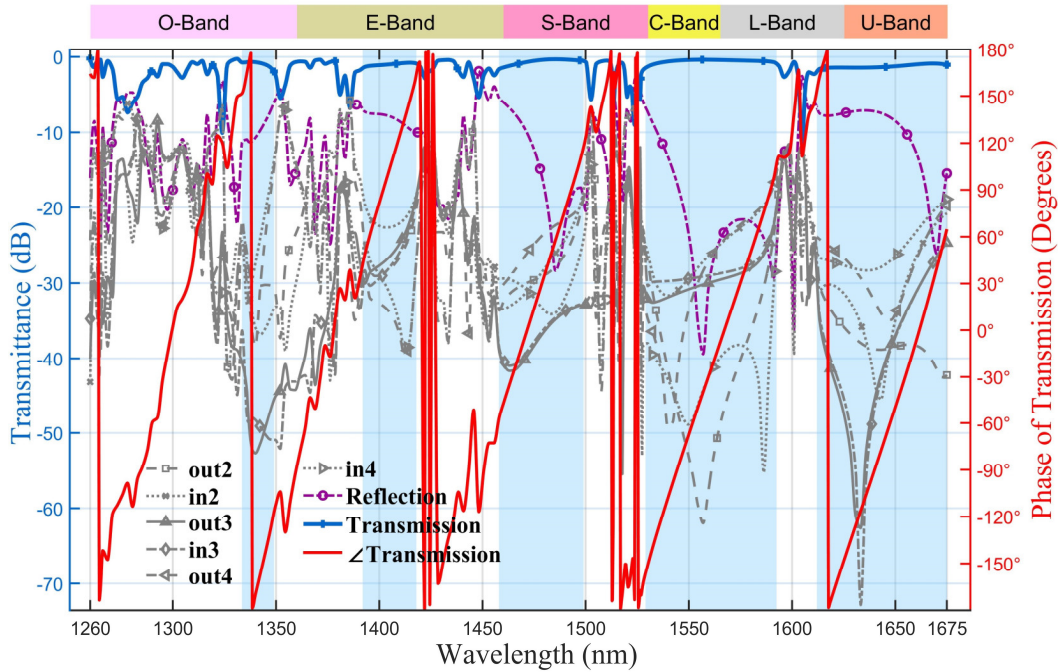

Fig. 6. Simulated transmission, reflection and crosstalk spectra of GPC-MFE based *4×4* waveguide intersection of figure 3(b). The wavelengths that the lens has the desired performance are highlighted in blue.

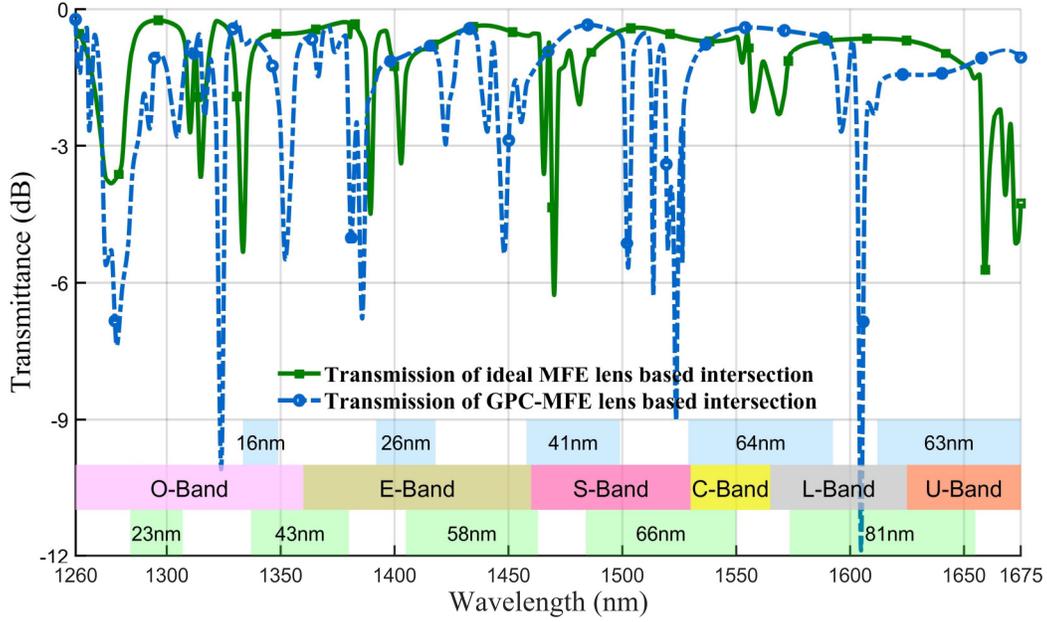

Fig. 7. Comparison of the ideal MFE and GPC-MFE lenses' transmission as *4×4* waveguide intersection. Green and blue horizontal bars specify the wavelengths with the acceptable transmission for ideal MFE and GPC-MFE lenses, respectively. The bandwidth of each transmission band is determined.

In order to prove that the MFE lens as intersection medium can support crossing of more waveguides, simulation results of a *6×6* waveguide intersection based on GPC-MFE lens with the radius of *10×a* are also presented. Due to the symmetry of the *6×6* waveguide intersection, there are two distinguishable electric field distributions as shown in Fig. 8(a) and (b). In the design of the *6×6* intersection, rods' radii are fixed to four different values and, consequently, the spacing between the rods is adjusted to implement the lens's refractive index profile. The *6×6* intersection is designed with the optimum sizes of rods to cover the entire C-band. As shown in Fig. 8(a), when an optical signal is applied to port *in5* at *1550nm*, return loss is *32.7dB* and insertion loss is *0.43dB* while crosstalk levels are below *-23.6dB*. And Fig. 8(b) shows the propagation of light at a wavelength of *1550nm* from port *in1* to *out1* through *6×6* intersection. At this wavelength, return loss of *39.2dB*, insertion loss of *0.43dB*, and crosstalk levels below *-23.6dB* are obtained. The simulated scattering parameters of GPC-MFE based *6×6* intersection of Fig. 8(a) and 8(b) are shown in Fig. 9 and 10 in the C-band, respectively. For the *6×6* intersection, the crosstalk levels are below *-18dB* and average insertion loss is *0.85dB* in the C-band.

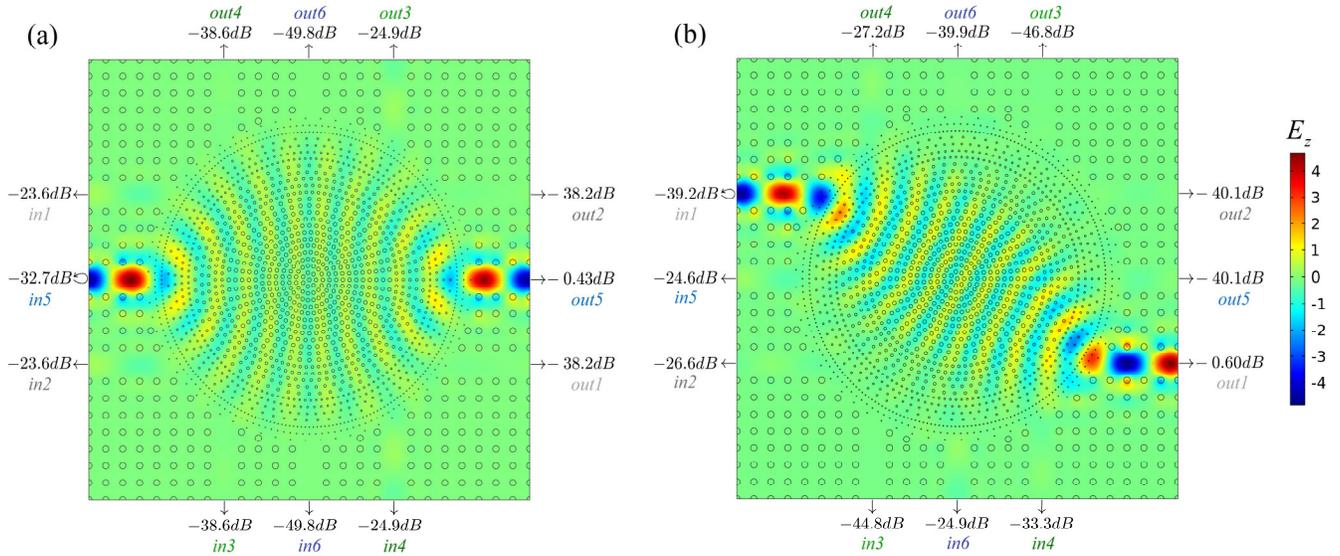

Fig. 8. Transmission, reflection, and crosstalk of *6×6* intersection based on GPC-MFE lens at a wavelength of *1550nm*. Due to the symmetry, (a) and (b) are the two possible electric field distributions.

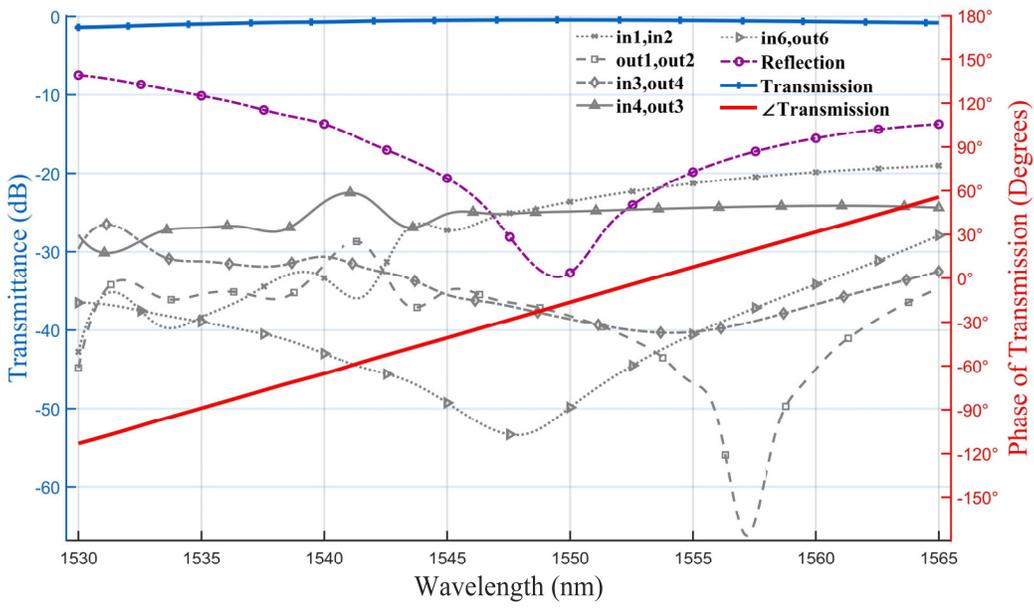

Fig. 9. Simulated transmission, reflection and crosstalk spectra of *6×6* intersection based on GPC-MFE lens of Fig. 8(a) in the C-band.

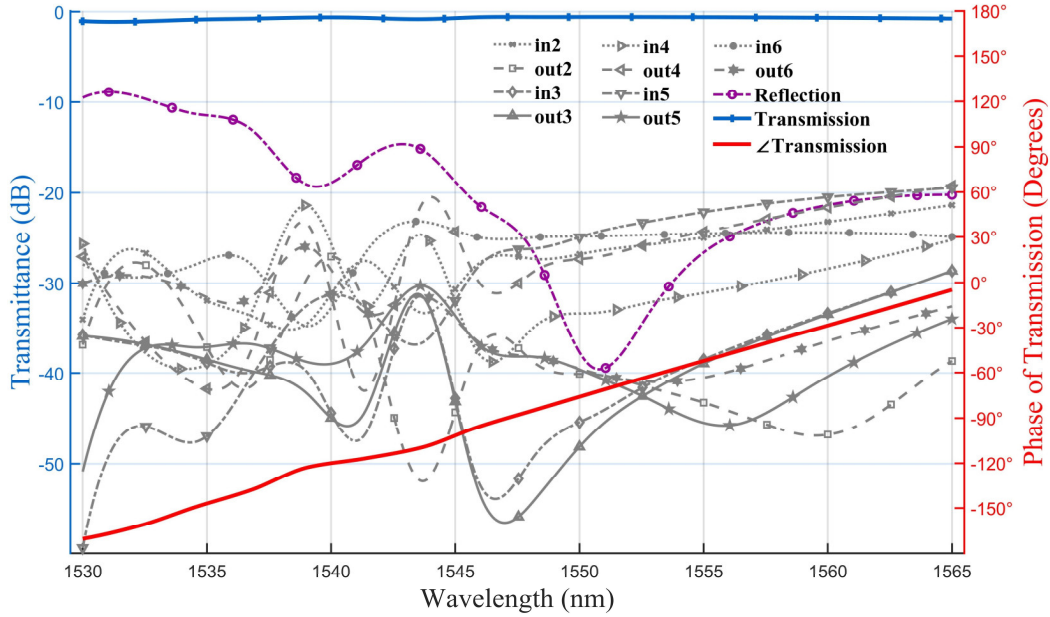

Fig. 10. Simulated transmission, reflection and crosstalk spectra of *6×6* intersection based on GPC-MFE lens of Fig. 8(b) in the C-band.

## 4. Conclusion

In this paper, it is shown that MFE lens can be used to cross multiple vertical and horizontal waveguides without any need to design bends. The optimized *4×4* and *6×6* intersections based on GPC-MFE lens are presented covering the entire C-band. Crosstalk levels of below *-24dB* and average insertion loss of *0.60dB* are achieved in the C-band for *4×4* intersection. Furthermore, for *6×6* intersection, crosstalk levels of below *-18dB* and average insertion loss of *0.85dB* are achieved in the C-band. The diameters of the proposed MFE lenses are *14×a* and *20×a* for *4×4* and *6×6* intersections, respectively. In the C-band, the transmission of proposed intersections is approximately constant with linear phase resulting in minimal dispersion and pulse broadening.